\documentclass[twocolumn,showpacs,preprintnumbers,amsmath,amssymb,pre]{revtex4}

\usepackage{graphicx}
\usepackage{subfigure}
\usepackage{dcolumn}
\usepackage{bm}

\begin{document}


\title{Dimensionality effects in dipolar fluids} 

\author{Remi Geiger$^1$ and Sabine H.~L.~Klapp$^2$}
\affiliation{$^{1}$Laboratoire Charles Fabry de l'Institut d'Optique, CNRS and Universit\'e Paris-Sud, RD 128, 91127 Palaiseau, France} 
\email{remi.geiger@u-psud.fr}
\affiliation{$^{2}$Institut f\"ur Theoretische Physik, Sekr. EW 7-1, Technische Universit\"at Berlin, Hardenbergstr. 36, 10623 Berlin, Germany}
  \email{klapp@physik.tu-berlin.de}


\date{\today}

\begin{abstract}
Using classical density functional theory (DFT) in a modified mean-field approximation we investigate 
the fluid phase behavior of quasi-two dimensional dipolar fluids confined to a plane. The particles carry three-dimensional dipole moments and interact via a combination
of hard-sphere, van-der-Waals, and dipolar interactions. The DFT predicts complex phase behavior involving
first- and second-order isotropic-to-ferroelectric transitions, where the ferroelectric
ordering is characterized by global polarization within the plane. We compare this phase behavior, particularly the onset of ferroelectric ordering and the related tricritical points, with 
corresponding three-dimensional systems, slab-like systems (with finite extension into the third direction), and true two-dimensional systems with two-dimensional dipole moments.
\end{abstract}

\verb + 

\maketitle



\section{Introduction}
\label{intro}
 Two-dimensional (2D) fluids consisting of particles with classical dipole-dipole interactions such as (para)magnetic nanoparticles at interfaces
 \cite{Butter03,Klok06,Ziherl07}, cobalt nanocrystals on solid surfaces \cite{Pileni08}, 
 and suspensions of polarizable colloids in 2D dielectrophoretic set-ups \cite{Velev04,Juarez09}, currently attract much attention. 
 Indeed, as a result of the directionality of the interactions whose details can be tuned
 by external (in-plane, out-of-plane, or tilted) fields, 2D dipolar systems display a variety of interesting structures such as chains and bundles at low densities \cite{Butter03,Klok06}
 but also various solid phases \cite{Ziherl07}. Especially the self-assembled low-density structures suggest that such systems are promising candidates as tunable advanced materials
 \cite{Glotzer07} with applications in electrical engineering, sensors \cite{Bhatt04} and molecular miniature devices. 
 
For theory and computer simulations, exploring the full structural and phase behavior of 2D dipolar systems remains challenging. Apart from the above-mentioned aggregation phenomena,
one topic investigated particularly by computer simulations concerns the appearance and characteristics of vapor-liquid transitions
\cite{Weis98,Tavares02,Tavares06,Duncan06,Gao97,Heiko}. Another question
touches the structure at high densities close to the range where crystallization is expected to occur. Various Monte Carlo (MC) simulation studies  \cite{Weis98,Lomba00} revealed the 
appearance of ferroelectric (or ferromagnetic, respectively) domains,
but overall frustrated (vortex) structures without true long-range orientational ordering. This is consistent with integral equation results \cite{Lomba00,Luo09},
where predictions on the low-temperature behavior are extracted by analyzing correlation functions. On the other hand, recent Molecular Dynamics (MD) simulations  
\cite{Ouyang11}
revealed long-range ferroelectric ordering in dense, 2D Stockmayer fluids, where the dipole-dipole interactions are supplemented by isotropic Lennard-Jones (LJ) interactions.

Similar to the dense fluid state, the nature
of the 2D {\em crystalline} structures formed at finite temperatures remains so far unclear 
\cite{weis_mono,russier}, although ground state calculations indicate ferromagnetism for certain 2D lattice types such as hexagonal lattices
\cite{politi06}. In three-dimensional (3D) systems 
the existence of long-range ferroelectric (-magnetic) ordering under appropriate boundary conditions
is well established \cite{Wei92,Weis92,Weis06}. Moreover, computer simulations of slab-like systems \cite{Klapp02}, where the particles are confined between two plane-parallel walls, have
indicated that this type of confinement can actually promote long-range ordering of the dipole moments, if the wall separation $L_{\text{z}}$ is sufficiently large. However, decreasing 
$L_{\text{z}}$ to values where less than three monolayers can form, the ordering seems to disappear \cite{Trasca08}.

The diversity of simulation results shows that spatial dimension has a profound influence of the ordering behavior of dipolar fluids. The purpose of the present study is to 
collect and compare theoretical results on that issue based on a relatively simple, mean-field like approach. Specifically, we employ classical density functional theory (DFT) in the
modified mean field approximation \cite{TEIX91,FROD92,TAVA95}, 
where the pair correlations are replaced by their low-density limit, i.e., the Boltzmann factor. The application of this approach for three-dimensional (3D)
dipolar systems and their mixtures \cite{Range04} was put forward by Groh and Dietrich \cite{Groh94PRL,Groh94PRE,Groh96PRE,Groh97PRL}, who considered Stockmayer fluids.
Later the modified mean-field DFT approach has been used to study confined, slab-like Stockmayer fluids  \cite{Gramzow07,Szalai09},
with different degrees of sophistication regarding the hard-sphere part of the density functional,
 
Here we apply the approach to a quasi-2D dipolar (Stockmayer) fluid, where the particles are confined to a plane, but carry 3D dipole moments. 
Evaluating the phase diagram
and comparing with corresponding DFT results  for 3D systems, slab-like systems, and true 2D systems with 2D dipole moments
we can identify, on a mean-field level, the influence of spatial dimension and of the dimension of the order parameter
on global ordering in fluid-like dipolar systems. Based on previous experiences one would expect that the mean-field approach for the quasi-2D system will (as it generally does)
overestimate the stability of orientationally ordered phases. However, given the importance of meanfield-like approaches in the general context of spin and dipolar systems, and realizing
that the mean-field DFT approach is, so far, still the only theory targeting the whole
(homogeneous) phase diagram of dipolar systems, we think our results are important for a complete understanding of such systems.

The remainder of the paper is organized as follows. In Sec.~\ref{theory} we formulate the quasi-2D model and briefly detail the derivation of the 
corresponding grand-canonical functional. Numerical results for the phase diagram at a typical dipole moment are presented in Sec.~\ref{results}. There we also use Landau expansions
to compare the onset of ordering in the quasi-2D system with the cases of 3D, slab-like
and true 2D systems. Finally, in Sec.~\ref{summary} we summarize our results.
\section{Theory}
\label{theory}
The quasi-2D Stockmayer fluid consists of disk-like particles of diameter $\sigma_{\text{T}}$
at positions $\mathbf{r}_i=(x_i,y_i)$ in the $x$-$y$ plane. The orientation of their 3D dipole moments $\hat{\boldsymbol{\mu}}_i$
is represented by the Euler angles $\omega_i=(\theta_i,\phi_i)$.
The microscopic interactions between the particles stem from anisotropic dipole-dipole and isotropic LJ forces. The resulting pair potential between two particles with coordinates $(1)=(\mathbf{r}_1,\omega_1)$ and $(2)=(\mathbf{r}_2,\omega_2)$ is given as
\begin{equation}
\label{eq:interactions}
u(1,2)= \left
\{
\begin{array}{ll}
\infty, \ r_{12} \leq \sigma_{\text{T}} \\
u_{\text{dip}}(\mathbf{r}_{12},\omega_1,\omega_2) + u_{\text{LJ}} (r_{12}), \ r_{12} > \sigma_{\text{T}},
\end{array}
\right.
\end{equation}
where $\mathbf{r}_{12}=\mathbf{r}_1 -\mathbf{r}_2$ is the connecting vector between the two particles, and $r_{12}=|\mathbf{r}_{12}|$.
Further, $u_{\text{dip}}(\mathbf{r}_{12},\omega_1,\omega_2)  =  (\mu^{2}/r_{12}^{3})\Big[ \hat{\boldsymbol{\mu}}_1(\omega_1) \cdot \hat{\boldsymbol{\mu}}_2(\omega_2)
 -  3(\hat{\boldsymbol{\mu}}_1(\omega_1) \cdot \hat{\mathbf{r}}_{12})(\hat{\boldsymbol{\mu}}_2(\omega_2) \cdot \hat{\mathbf{r}}_{12}) \Big]$ is the 3D 
 dipole-dipole interaction potential,
and $u_{\text{LJ}}(r_{12})=4 \epsilon \Big[ \Big(\sigma/r_{12}\Big)^{12} - \Big(\sigma/r_{12}\Big)^{6} \Big]$ is the LJ potential.
 To mimic the fact that the effective range of the $r^{-12}$ repulsion varies with the thermodynamical parameters, we choose in our DFT calculations
 a temperature-dependent hard core defined via the Barker-Henderson formula \cite{Barker67}, that is, $\sigma_{\text{T}}=\int_{0}^{\sigma} dr\left(1-\exp\left[-\beta u_{\text{LJ}}(r)\right]\right)$, where
 $\beta=1/k_{\mathrm{B}}T$ (with $k_{\mathrm{B}}$ and $T$ being the Boltzmann constant and the temperature, respectively).

To analyze the phase behavior we employ classical DFT, where the key quantity is the grand canonical potential $\Omega$ as a functional of the singlet density  $\rho(\mathbf{r},\omega)=\Big\langle \sum_{i=1}^{N}\delta(\mathbf{r}-\mathbf{r}_i)\delta(\omega-\omega_i) \Big\rangle$, with $N$ being the total number of particles \cite{McDonald}.  
We restrict the analysis to fluid-like but possibly {\em polarized} ordered phases of the quasi-2D Stockmayer fluid (in the following we assume, without loss of generality, the dipoles to be
of electric nature).
In principle, investigation of this situation requires to perform a {\em free} minimization for the profile $\rho(\mathbf{r},\omega)$, 
thereby allowing the system to form domains (or other patterns with spatially varying polarization). Practically, however, minimization including pattern formation is a quite challenging task as demonstrated in \cite{Groh97PRL,Groh96PRE}.
In the present study, where we are interested in the general tendency for ordering, we neglect that problem.
That is, we focus on the polarization {\em within} a (macroscopically large) fluid domain, which {\em may} be part of a globally unpolarized system.
We thus consider the singlet density $\rho(\mathbf{r},\omega)=\rho \alpha(\omega)$, with $\rho$ being the constant number density of particles and $\alpha(\omega)$ the orientational distribution function of their dipole moments. 
This function is normalized to $1$ (i.e., $\int{ \alpha(\omega) d \omega} =1$) and equals $1/4\pi$ for isotropic states \cite{Gramzow07}. 
To describe an orientational ordering along a specific direction,
 we expand the orientational distribution function in terms of spherical harmonics, that is, 
 $\alpha(\omega)=\sum_{l=0}^{\infty}\sum_{m=-l}^{l}\alpha_{lm}Y_{lm}(\omega)$, with the coefficients $\alpha_{lm}$ representing orientational order parameters \cite{Gramzow07}. 
 Nonzero order parameters with $l=1$ correspond to a macroscopic polarization $\mathbf{P}=\rho\mu\mathbf{p}$, 
 with $\mathbf{p}=\int\alpha(\omega)\hat{\boldsymbol{\mu}}(\omega)d \omega$. 
 Indeed, transforming the Cartesian components of $\mathbf{p} = \left(p_{\mathrm{x}},p_{\mathrm{y}},p_{\mathrm{z}}\right)$ 
in terms of spherical harmonics one obtains 
$\mathbf{p} =  \sqrt{4\pi/3} \ \left(-2\text{Re}(\alpha_{11}),2 \text{Im}(\alpha_{11}),\alpha_{10}\right)$ \cite{Gramzow07}. Similarly,
nonzero order parameters $\alpha_{lm}$ with $l=2$ indicate that there is a preferred orientation of the dipole axes. 

Within the grand canonical formalism, the system is characterized by its size $\mathcal{A}$, the chemical potential $\mu_{\text{chem}}$ and the inverse temperature 
$\beta$. In the present study, the dipolar contribution to the excess (interaction) free energy is treated in the modified mean-field approximation, where the pair correlations are approximated
by the Boltzmann factor \cite{TEIX91,FROD92,TAVA95}. In addition, following our previous study on slab-like systems \cite{Gramzow07},
we perform a perturbation expansion of the Mayer function and truncate the latter after the quadratic term. Such a truncation has been first employed in a
DFT study of the surface tension of polar fluids \cite{TEIX91}. Later,
results for the phase diagrams of bulk polar fluids
\cite{Groh94PRE} have indicated
that the second--order theory yields data very close to those from the full modified meanfield approximation 
(without any truncation). As a consequence of the truncation, the resulting excess free energy contains only terms up
to $l=2$. For a detailed calculation for the quasi-2D case (which proceeds analogous to the slab case) we refer the reader to Ref.~\cite{GeigerThesis}. 
The resulting expression for the grand canonical functional is given by
\begin{eqnarray}
\label{omegafinal}
\frac{\beta\Omega}{\mathcal{A}} &  =  & \rho \left(\ln\left(\rho\lambda^{2}\right)-1 -\beta\mu_{\text{chem}}+\int d\omega \alpha(\omega) \ln\left[4\pi\alpha(\omega)\right] \right)\nonumber \\ 
 & - & \rho\ln(1-\eta_T) +\rho\frac{\eta_T}{1-\eta_T} \nonumber \\
 & - & \pi\rho^2\sigma^2 g_1(T) \nonumber \\
 & + & \frac{4\pi^2}{3}\mu^2\rho^2\beta\big(\sigma_T^{-1}+\sigma^{-1} g_2(T)\big) \big( \alpha_{10}^2 - |\alpha_{11}|^2 \big) \nonumber \\
 & - & \mu^4\rho^2\beta^2\sigma^{-4} g_3(T) G[\alpha(\omega)]. \nonumber \\
\end{eqnarray}
On the right side of Eq.~(\ref{omegafinal}), the first line contains ideal gas contributions (involving the thermal wavelength $\lambda=h/\sqrt{2\pi\beta m}$) and
the orientational entropy (last term). The second line contains the excess free energy of our reference system, the hard disk fluid \cite{Nielaba97}, 
involving the 2D packing fraction $\eta_T=\pi\rho\sigma_T^2/4$. 
The three last terms account for the dipolar interactions, where the functions $ g_1(T)=\int_{a(T)}^{\infty}  dx\,x \left(\exp\left[-4\beta\epsilon(x^{-12}-x^{-6})\right]-1\right)$, 
$g_2(T) =  \int_{a(T)}^{\infty} dx\,x^{-2}\left(\exp\left[-4\beta\epsilon(x^{-12}-x^{-6})\right]-1\right)$, 
and $g_3(T) = \int_{a(T)}^{\infty} dx\,x^{-5}\left(\exp\left[-4\beta\epsilon(x^{-12}-x^{-6})\right]\right)$, with $a(T)=\sigma_{\text{T}}/\sigma$.
Finally, in the last term on the right side of Eq.~(\ref{omegafinal}), $G[\alpha(\omega)]  = t_1\alpha_{00}^2 +t_2\alpha_{00}\alpha_{20} +t_3\alpha_{20}^2 +t_4|\alpha_{21}|^2 +t_5|\alpha_{22}|^2$, where $t_1=10\pi^2/(15\sqrt{2})$, $t_2=-2\sqrt{5}\pi^2/(15\sqrt{2})$, $t_3=3\pi^2/(15\sqrt{2})$, $t_4=-2\pi^2/(15\sqrt{2})$ and  $t_5=\pi^2/(15\sqrt{2})$.

Minimization of the functional \eqref{omegafinal} with respect to $\rho$ and $\alpha(\omega)$ yields the Euler-Lagrange equations for this problem \cite{GeigerThesis}. 
They consist of a set of nonlinear, coupled equations for the density $\rho$ and the OPs $\alpha_{lm}$ appearing in Eq.~(\ref{omegafinal}). The
equations are solved numerically using a Newton-Raphson algorithm \cite{NumRec}. 

In the following we characterize the state of the quasi-2D Stockmayer fluid by the dimensionless density $\rho^*=\rho\sigma^2$, 
temperature $T^*=k_{\mathrm{B}}T/\epsilon$, chemical potential $\mu_{\text{chem}}^*=\epsilon^{-1}(\mu_{\text{chem}} -2k_{\mathrm{B}}T\ln(\lambda/\sigma))$,
and dipole moment $m^*=\mu/\sqrt{\epsilon \sigma^3}$. The quantity $m^{*2}$ measures the strength of the dipolar interactions in an antiparallel side-by-side configuration relative to the LJ interactions. We note that the coupling parameters $m^*$ and $T^*$ are equivalently defined in 3D (or slit-pore-) dipolar systems \cite{Gramzow07}, so that the
results can be conveniently compared. 

\section{Results}
\label{results}
\begin{figure}
\includegraphics[width=8cm]{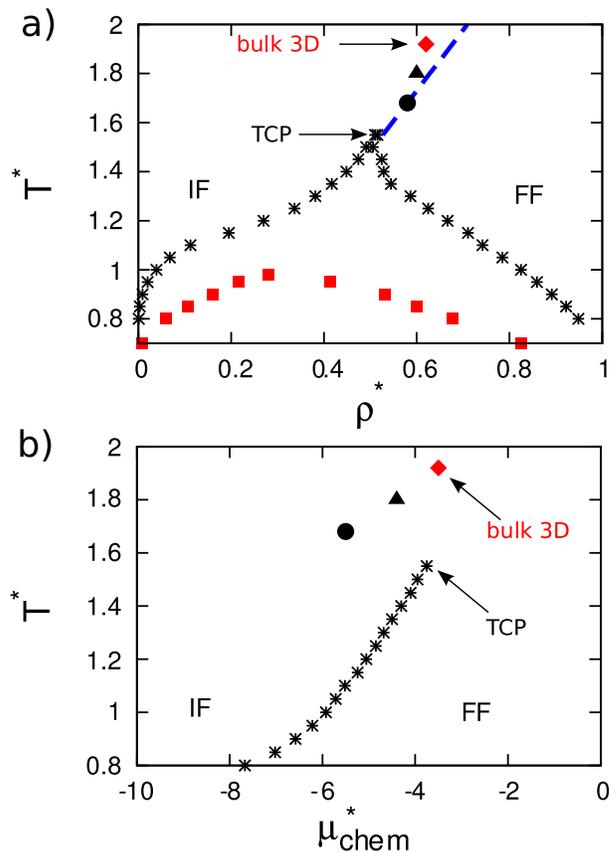}
\caption{(Color online) a) Phase diagram of the quasi-2D Stockmayer fluid  ($m^*=1.5$) in the density-temperature plane. The regime under the black stars indicates the IF-FF coexistence region. The blue dashed line is the line of critical points [see Eq.~\eqref{Tc_IF_FF}], which starts at a tricritical point (TCP). Included are results (from \cite{Gramzow07})
for the TCPs of the corresponding 3D Stockmayer fluid (red diamonds), and two slit-pore systems with wall separations $L_{\mathrm{z}}=10\sigma$ 
(black triangle) and $L_{\mathrm{z}}=4\sigma$ (black circle). The red squares indicate the IG-IL coexistence curve of the quasi-2D system
resulting from a separat calculation where the system is forced to be unpolarized. b) Corresponding phase diagram in the chemical potential-temperature plane.}
\label{Stock_m15_T}
\end{figure}

Following earlier DFT studies on confined Stockmayer fluids \cite{Gramzow07} we consider a system characterized by $m^* = 1.5$,
a typical value for moderately polar molecular fluids such as chloroform \cite{Leeuwen}. The calculated fluid phase diagram in the density-temperature and chemical potential-temperature plane 
is shown in Figs.~\ref{Stock_m15_T}a) and \ref{Stock_m15_T}b), respectively. The latter representation
better relates to typical sorption experiments \cite{Schreiber02}. 
For small and intermediate densities (or chemical potentials) we find a state
where all order parameters $\alpha_{1m}$ are equal to zero, and those with $l=2$ are either zero ($m=\pm 1,\pm 2$) or negative ($m=0$). Thus, there is no global polarization and neither a global ordering of the dipole axes; we therefore refer to this state as "isotropic fluid" (IF). The negative values
of $\alpha_{20}$ merely indicate that the dipoles tend to avoid to be oriented parallel or antiparallel to the $z$-axis; rather they prefer to lie (with random orientations) in the $x$-$y$-plane. This is an expected effect in a dilute, quasi-2D dipolar system (consistent with simulations and other theoretical studies, see e.g. \cite{Luo09}).
At higher densities, the system then develops a non-zero polarization, to which we refer to as "ferroelectric fluid" (FF). Within this state, the vector
$\mathbf{P}$ points along an (arbitrary) direction in the $x$-$y$-plane (as reflected by $\alpha_{1,\pm 1}\neq 0$, $\alpha_{10}=0$). 
Notice that the preference of in-plane polarization (rather than out-of-plane polarization) can already be seen from the prefactor of the corresponding terms 
$\propto |\alpha_{1m}|^2$ in Eq.~(\ref{omegafinal}). 

The transition between the IF and the FF phase is discontinuous in $\alpha_{lm}$ and $\rho$ (yet not in $\mu_{\text{chem}}$)
for temperatures below a tricritical temperature $T^*_{\text{tcp}}=1.57$ 
[see Fig.~\ref{Stock_m15_T}a)]. 
Above the tricritical point (TCP) the transition becomes continuous, which results in a line of critical points. The appearance of a TCP is a typical feature
of DFT predictions of the phase diagrams of dipolar fluids \cite{Gramzow07,Groh94PRE,Groh94PRL} and also Heisenberg fluids \cite{Lomba94}. 
Recent MC studies for 3D dipolar fluids confirm that the transition between isotropic and ferroelectric fluid is of second-order
in a broad range of temperatures \cite{Weis06}.
Within the DFT, the line of critical points can be determined from a Landau expansion
of the free energy, assuming that the OPs characterizing the FF state are small
(i.e., $|\alpha_{1m}|\ll\alpha_{00}=1/\sqrt{4\pi}$). To this end we expand the integral $I[\alpha(\omega)]=\int d\omega \alpha(\omega) \ln\left[4\pi\alpha(\omega)\right]$ in Eq.~\eqref{omegafinal}, that is, the orientational entropy, in a Taylor serios around the isotropic state (where $\alpha(\omega)=\alpha_{00}Y_{00}(\omega)=1/4\pi$).
Collecting those terms in the resulting approximate functional, $\tilde{\Omega}$, that are proportional to $\alpha_{1,\pm1}$, we obtain  \cite{GeigerThesis}
\begin{equation}
\label{omega_propto_a11}
\frac{\beta\tilde{\Omega}}{\mathcal{A}}\propto |\alpha_{11}|^2 \big( 4\pi\rho-\frac{4\pi^2}{3}\mu^2\rho^2\beta\big(\sigma_T^{-1}+\sigma^{-1} g_2(T)\big)   \big),
\end{equation}
where the first term stems from the orientational entropy, whereas the second term results from the interaction free energy in Eq.~(\ref{omegafinal}). 
The second order phase transition is characterized by a change of sign of the factor of $|\alpha_{11}|^2$ in Eq.~(\ref{omega_propto_a11}). We thus obtain
\begin{equation}
\label{Tc_IF_FF}
k_{\text{B}} T_{\mathrm{c}}^{\text{quasi-2D}}=\frac{\pi}{3}\mu^2\rho\big(\sigma_{T}^{-1}+\sigma^{-1} g_2(T_{\mathrm{c}})\big).
\end{equation}
This is approximatively the equation of a straight line in the  density-temperature plane, consistent with what one sees in Fig.~\ref{Stock_m15_T}a).

In Figs.~\ref{Stock_m15_T}a) and b) we have included DFT data for tricritical points of Stockmayer fluids in 3D and in slit-pore geometries. Within the latter situation,
the particles are confined between two planar, attractive walls separated by a distance $L_{\mathrm{z}}$ \cite{Gramzow07}. We note that, both for the 3D and the slab case, our data somewhat differ numerically from those presented in another recent DFT study
of confined Stockmayer fluids \cite{Szalai09}. This is 
since we used (contrary to \cite{Szalai09}) a temperature-dependent particle diameter and a homogeneous ansatz for the number density in the slit-pore. However, from a qualitative point of view, the observed trends regarding the impact of confinement on the TCP are the same in both studies \cite{Gramzow07,Szalai09}. 
In particular, both predict that decreasing $L_{\mathrm{z}}$ (that is, increasing the degree of
confinement) shifts the TCP towards  lower temperatures and somewhat lower densities
For example, according to \cite{Gramzow07}, the tricritical parameters $(\rho^{*}_{\mathrm{tcp}},T^{*}_{\mathrm{tcp}})$
are $(0.62,1.92)$ at $L_{\mathrm{z}}=\infty$, i.e., in the 3D (bulk) limit, $(0.60,1.80)$ at $L_{\mathrm{z}}=10\sigma$ and $(0.58,1.68)$ at $L_{\mathrm{z}}=4\sigma$ \cite{Gramzow07}. Consistent
with this tendency,  the TCP of the quasi-2D Stockmayer fluid 
(which corresponds to the limit $L_{\text{z}}\rightarrow 0$) is found at even lower density and temperature, specifically 
at $\rho^*_{\text{tcp}}= 0.52$ and $T^*_{\text{tcp}}= 1.57$. 
A somewhat different behavior emerges in the chemical potential-temperature representation depicted in Fig.~\ref{Stock_m15_T}b). From the location of the TCPs, we see
that  the tricritical chemical potential, $\mu_{\text{chem}}^{\text{tcp}}$, of the quasi-2D fluid ($L_{\mathrm{z}}=0$)
is only slightly smaller than that of its 3D counterpart ($L_{\mathrm{z}}=\infty$). On the other hand, the corresponding values for $\mu_{\text{chem}}^{\text{tcp}}$
of confined Stockmayer fluids ($L_{\mathrm{z}}=10\sigma$, $4\sigma$)
are significantly smaller (consistent with \cite{Szalai09}). We attribute this non-monotonic behavior of $\mu_{\text{chem}}^{\text{tcp}}$ upon lowering of $L_{\text{z}}$
[see Fig.~\ref{Stock_m15_T}b)]
to the fact that, for the confined Stockmayer fluids, the walls were considered to be attractive \cite{Gramzow07}. This feature is known to support capillary condensation (or, more generally,
the formation of denser phases), accompanied by lowering of the critical value of $\mu_{\text{chem}}$. The particles in the quasi-2D system do not interact with any walls; thus, there is no capillary condensation phenomenon. As a result, $\mu_{\text{chem}}^{\text{tcp}}$ for the quasi-2D system nearly agrees with the 3D value.

The reduction of spatial dimension not only {\em shifts} the TCP, it also has a profound influence on the {\em topology} of the phase diagram. 
Indeed,  while the 3D Stockmayer fluid with the same dipole moment ($m^{*}=1.5$) exhibits, in addition to the IF-FF transition, a condensation transition 
within the isotropic liquid (IL) phase, such a transition is absent in the quasi-2D system (see Fig.~\ref{Stock_m15_T}). We can artificially stabilize a condensation transition
in the 2D system by setting all order parameters (except from $\alpha_{20}$) to zero. The result of this calculation is indicated in Fig.~\ref{Stock_m15_T}a) by the red squares. It turns out that the IG-IL critical point ($T^*_{\mathrm{c}}=0.98$) is located within the IF-FF phase coexistence region; therefore it is thermodynamically unstable. 
However, such IL configurations may still be relevant in the context of the phase separation kinetics (i.e., in non-equilibrium situations), where they can occur
as transient states during the change from the IF to the FF state at a temperature $T<T^*_{\mathrm{c}}$. We also note that
the suppression of the IG-IL critical point is consistent 
with previous DFT results (at $m^{*}=1.5$)
in very narrow slit-pores, such as $L_{\mathrm{z}}=4\sigma$ \cite{Gramzow07}. On the other hand, MC simulations for quasi-2D Stockmayer fluids
predict stable isotropic liquid phases for dipole moments up to (at least) $m^{*}=\sqrt{6}$ \cite{Gao97,Heiko,Ouyang11}. Therefore, the DFT seems to overestimate the stability
of the ferroelectric phase, similar as it does in 3D \cite{Gramzow07}. Within the DFT, one would expect a recovery of the isotropic liquid state when the LJ attraction finally dominates then dipolar coupling, i.e., when $m^{*}$ decreases towards even smaller values.

\begin{figure}[!h]
\includegraphics[width=8cm]{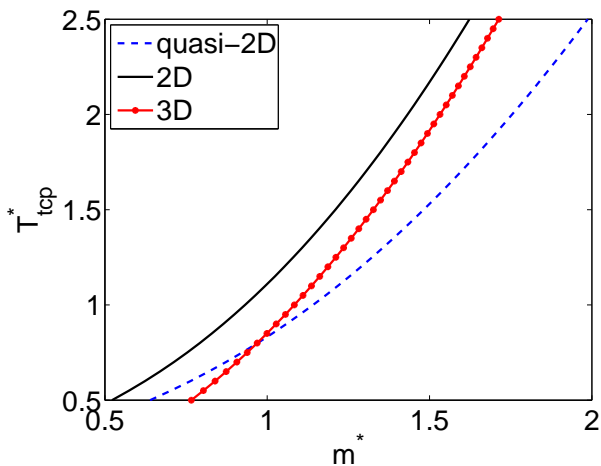}
\caption{(Color online) Tricritical temperature $T_{\text{tcp}}^*$ as function of the dipole moment $m^*$ in the quasi-2D, 3D, and true 2D geometry. In all cases, the density has been set
to that obtained at $m^{*}=1.5$.}
\label{T_tcp_m}
\end{figure}

We now discuss in more detail the influence of spatial dimension and its interplay with the dipolar coupling strength
 on the tricritical point $(\rho^{*}_{\mathrm{tcp}},T^{*}_{\mathrm{tcp}})$, 
 above of which the low-temperature discontinuous transition between the IF and FF states changes into a (line of) second-order transition(s). 
 Specifically, we are interested in the position of the TCPs, in the quasi-2D system and its 3D counterpart, 
as functions of the parameter $m^{*}$. In previous DFT
studies \cite{Gramzow07,Groh94PRE,Groh94PRL} it was already shown that the coupling strength influences the quantity $T^*_{\text{tcp}}$ 
much more than $\rho^*_{\text{tcp}}$, at least as long as $m^* \gtrsim 1.0$. Therefore, to estimate the dependence of $T^*_{\text{tcp}}$ on $m^{*}$ in the quasi-2D system, 
we {\em set}
the density equal to tricritical density at $m^*=1.5$, that is, to $\rho^{*}_{\text{tcp}}|_{\text{quasi-2D}}= 0.52$. The resulting function $T^*_{\text{tcp}}\left(m^{*}\right)$ can then be easily determined
from Eq.~(\ref{Tc_IF_FF}). The same procedure is used for the 3D case,
where the analog of Eq.~(\ref{omega_propto_a11}) reads \cite{note}
\begin{equation}
\label{omega_3D}
\frac{\beta{\tilde{\Omega}}^{\text{3D}}}{\mathcal{V}} \propto |\alpha_{11}|^2 \big( 4\pi\rho_{\text{b}}-\frac{16\pi^2}{9}\beta\mu^2\rho_{\text{b}}^2 \big)
\end{equation}
with $\mathcal{V}$ being the volume and $\rho_{\text{b}}=\rho^*\sigma^3$ being the density of the bulk system. 
Equation~(\ref{omega_3D}) yields $k_{\text{B}}T_{\mathrm{c}}^{\text{3D}}=(4\pi/9)\rho_{\text{b}}\mu^2$. Fixing the density to that of the tricritical point at $m^{*}=1.5$ \cite{Gramzow07},
$\rho^{*}_{\text{tcp}}|_{3D}= 0.61$, we can again estimate the function $T^*_{\text{tcp}}\left(m^{*}\right)$ for a range of dipole moments.
Numerical results for the quasi-2D system and its 3D analog are plotted in Fig.~\ref{T_tcp_m}.
For both systems, the tricritical temperature increases with $m^{*}$, as one may expect when the dipolar interactions (which stabilize the FF state)
become more and more important as compared to the spherical attractive ones. More interestingly, Fig.~\ref{T_tcp_m} reveals that reduction of spatial dimension shifts
the tricritical temperatures towards {\em lower} values at fixed $m^{*}$. This shift can be reasoned from Eqs.~(\ref{omega_propto_a11}) and (\ref{omega_3D}): in both the quasi-2D and the 3D
system, ordering competes
with the same amount of (orientational) entropy, but the associated decrease of interaction energy ($\propto \rho^2\mu^2$) is less pronounced in the quasi-2D system. From a physical point of view, this 
diminishment is a consequence of the reduction of the number of neighbors in a 2D system relative to the bulk case. 
We also note another interesting point: whereas in the 3D system, only the long-range dipolar interactions contribute to the onset of ordering [see Eq.~(\ref{omega_3D})],
the corresponding onset in the quasi-2D system is also affected by the short-range interactions, as reflected by the appearance of the function $g_2(T)$ in Eq.~(\ref{omega_propto_a11}).
Since the $g_2$ function increases upon cooling down, the 
curves $T^*_{\text{tcp}}\left(m^{*}\right)$ in Fig.~\ref{T_tcp_m} for the quasi-2D and the 3D system, respectively, cross each other near $m^*=1$.
However, since we assumed a {\em constant} tricritical density, the precise position of this crossing in Fig.~\ref{T_tcp_m} should be considered with some caution.

Finally, we briefly discuss the influence of the dimension of the {\em dipole vector} (rather than that of the space accessible for the particles) on the appearance of ferroelectric order. 
Specifically, we consider a "true" 2D system where, in addition to the spatial confinement of the particles within the $x$-$y$-plane, the orientations of the dipole vectors are restricted to that plane as well.
Indeed, as shown in previous simulations and theoretical studies (see, e.g., \cite{Lomba00}) of 2D systems with freely rotating (i.e., three-dimensional) dipoles, these have a strong tendency to tilt into the confining plane especially at large coupling strengths. The "true" 2D system is therefore not completely unphysical.
In the true 2D case, the orientational distribution function $\alpha(\omega)$ depends only on one angle, $\phi$,
 which describes the orientation of $\hat{\boldsymbol{\mu}}_i$ relative to, say, the $x$-axis. To obtain the grand canonical functional we expand $\alpha(\phi)$ in a basis of exponential
 functions \cite{Luo09}, i.e., $\alpha(\phi)\propto\sum_{m}\alpha_m\exp\left[i m \phi\right]$. Performing then the same Landau expansion as for the quasi-2D system, and isolating the terms proportional to the polarization, i.e. to $\alpha_{m=\pm 1}$, the analog of Eq.~(\ref{omega_propto_a11}) reads
\begin{equation}
\frac{{\tilde{\Omega}}^{\text{2D}}}{\mathcal{A}}\propto|\alpha_1|^2 \big(2\pi\rho-\pi^2\mu^2\rho^2\beta \big(\sigma_{T}^{-1}+\sigma^{-1} g_2(T)\big)\big).
\label{energy_pure_2D}
\end{equation}
A direct comparison of Eq.~(\ref{energy_pure_2D}) with its quasi-2D analog in Eq.~(\ref{omega_propto_a11}) shows that, at fixed density, 
the ferroelectric ordering in the true 2D system occurs at a higher temperature. This is a consequence of the decrease of the dipolar fluctuations (and thus, the orientational entropy) 
due to their restriction to the plane.
Moreover,  as revealed in Fig.~\ref{T_tcp_m} by the corresponding curve $T^*_{\text{tcp}}\left(m^{*}\right)$, the ordering is even promoted relative to the 3D case. This is consistent
with tendencies found in a recent integral equation study \cite{Luo09}, although the latter predicts, for the low-temperature behavior,
large ferroelectric {\em domains} rather than true long-range ferroelectric order.

\section{Summary and conclusions}
\label{summary}
In this work we have calculated the fluid phase diagram of a quasi-2D Stockmayer fluid by means of density functional theory in the modified mean-field approximation. At the dipole moment considered
($m^{*}=1.5$) the system exhibits an isotropic fluid phase where the dipole moments are randomly oriented, yet with a preference for in-plane directions,
and a ferroelectric fluid phase characterized by global, in-plane polarization. Apart from exploring the quasi-2D phase behavior, another focus of our study
 was to identify the role of the dimension of accessible space, as well as that of the dimension of the dipole vector. To quantify these effects on a mean-field level, we have considered the location
 of the tricritical point. Regarding the impact of space dimension, we have found that decreasing the system's dimension in $z$-direction from the bulk limit
 ($L_{\text{z}}\rightarrow\infty$) over slab systems ($L_{\text{z}}=10\sigma$, $4\sigma$) 
 towards the 2D limit ($L_{\text{z}}\rightarrow 0$) shifts the TCP towards lower temperatures and densities. Furthermore,
 the disappearance of the isotropic liquid phase in the quasi-2D system also shows that the confinement enhances the stability of dense ordered phases relative to disordered ones.
 Clearly, care has to be taken with respect to the predictions of our mean-field-like DFT approach on a quantitative level. Indeed, from computer simulations \cite{Gao97,Heiko,Ouyang11} it is known
 that a quasi-2D Stockmayer fluid at $m^{*}=1.5$ does have a stable isotropic liquid phase at densities beyond the isotropic vapor-liquid critical point, which is absent in our study. 
 This discrepancy reflects the well-known tendency of the DFT
 to overestimate the stability of ordered phases. However, based on 
 previous DFT studies for bulk  and confined systems one would expect a recovery of the isotropic liquid state in the quasi-2D case upon further decrease of $m^{*}$. A further interesting
 result of our study concerns the role of the spin dimension. Here we have found that complete restriction of the dipole moments on in-plane directions yields ferroelectric ordering
 at temperatures not only higher than those in the quasi-2D system, but even higher than those in 3D. 
 
There remains the question whether fluid states with {\em long-range} ferroelectric order, as predicted by DFT, exist {\em at all} in quasi-2D and true 2D systems. As mentioned in the introduction, computer simulations give conflicting answers, which may also depend on the number of particles considered in the simulation (indeed, the MD study on quasi-2D systems by Ouyang {\em et al.} \cite{Ouyang11}, which does predict long-range ferroelectric ordering, involves a rather small system size). We should therefore interpret the present DFT results,
which rely on the assumption of a spatially homogeneous orientational structure, 
such that the 2D geometry definitely promotes the existence of large ferroelectric domains, but not necessarily true long-range order. 

Despite these pitfalls, the DFT approach provides a general overview of the phase diagrams and highlights the dimensionality effects by providing the leading order terms in the free energy. From that perspective, it would be interesting to extend the study towards 2D systems in external fields.



\begin{thebibliography}{99}
\bibitem{note} To obtain Eq.~(\ref{omega_3D}) we have taken the limit $L_{\text{z}}/\sigma_{\text{T}}\rightarrow\infty$
of the corresponding equation for slab-like systems (see Eq.~(2.53) in \cite{Gramzow07}).
\end{thebibliography}
\end{document}